\documentclass[manuscript]{aastex}

\newcommand{\myemail}{mtakuma@kwasan.kyoto-u.ac.jp}

\shorttitle{Alfv\'{e}n wave driven by photospheric motions}
\shortauthors{\textsc{MATSUMOTO} and \textsc{SHIBATA}}

\begin{document}


\title{Nonlinear propagation of Alfv\'{e}n waves driven by observed photospheric motions \\
: Application to the coronal heating and spicule formation}


\author{Takuma \textsc{MATSUMOTO} and Kazunari \textsc{SHIBATA}}
\affil{Kwasan and Hida Observatories, Kyoto University
   , Yamashina-ku, Kyoto, 607-8417, Japan; \myemail}
   


\begin{abstract}
We have performed MHD simulations of Alfv\'{e}n wave propagation along an open flux tube in the solar atmosphere.
In our numerical model, Alfv\'{e}n waves are generated by the photospheric granular motion.
As the wave generator, we used a derived temporal spectrum of the photospheric granular motion from \textit{G}-band movies of Hinode/SOT.
It is shown that the total energy flux at the corona becomes larger and the transition region height becomes higher in the case when we use the observed spectrum rather than white/pink noise spectrum as the wave generator.
This difference can be explained by the Alfv\'{e}n wave resonance between the photosphere and the transition region.
After performing Fourier analysis on our numerical results, we have found that the region between the photosphere and the transition region becomes an Alfv\'{e}n wave resonant cavity. 
We have confirmed that there are at least three resonant frequencies, 1, 3 and 5 mHz, in our numerical model.
Alfv\'{e}n wave resonance is one of the most effective mechanisms to explain the dynamics of the spicules and the sufficient energy flux to heat the corona.

\end{abstract}





\section{Introduction}

From the time the solar corona was found to be at a temperature of more than 1 million K \citep{edle43}, 
the search for the process that heats the corona, so called coronal heating problem, has become one 
of the most important problems in the solar physics.
It is known that mechanical energy transport is required to heat the corona because the photospheric temperature is at most 6,000 K.
Since large amount of mechanical energy flux can be generated by the photospheric convection motion, 
the essential problem is how to transport the mechanical energy to the corona and how to dissipate it in the corona.
The mechanisms that heat the actual corona are yet to be elucidated,
since it is difficult to self-consistently include all the energy transport process.

It is required that the heating rate of any coronal heating mechanisms must balance the energy loss by radiative cooling and thermal conduction
in order to maintain the 1 million K corona. The energy flux necessary to heat the quiet corona is $3\times10^{5}$ erg cm$^{-2}$ s$^{-1}$ \citep{with77}.
There are two mechanisms that are widely believed to heat the corona.
There are the Alfv\'{e}n wave heating model \citep{uchi74,went74}, and the the nanoflare model \citep{park88}.
Both models explain how to release Poynting flux generated by the photospheric convection motion.

In the Alfv\'{e}n wave heating model, 
magnetoacoustic waves driven by the convection motion propagate upward to produce shocks.
Damping length of compressional waves such as fast or slow mode waves is very short because of the highly stratified atmosphere in the Sun.
Moreover the diffraction due to the gradient of Alfv\'{e}n speed prevents compressional waves from propagating to the corona.
Therefore it is considered that it is not possible to heat the corona by these compressional waves \citep{holl81}.
On the other hand, Alfv\'{e}n waves are non-compressional waves, so they can reach the corona even if the atmosphere is highly stratified.
Damping length of Alfv\'{e}n modes is defined by various dissipation processes such as nonlinear mode conversion \citep{holl82b,kudo99,suzu05,suzu06} ,
Alfv\'{e}n wave resonance \citep{ions78}, and phase mixing \citep{heyv83}.
Alfv\'{e}n wave resonance also occurs when a 
flux tube between the photosphere and the chromosphere acts as an Alfv\'{e}n wave resonant cavity, 
this can happen even if it is not closed \citep{holl81,lero81}.
Since a large amount of the wave flux is reflected by the transition region, standing waves appear in the region between the photosphere and the transition region.
These standing waves can be amplified depending on the wave frequency excited at the wave source and they can eventually be dissipated nonlinearly. 
It is not understood yet which processes can act efficiently in the real solar corona.

Various studies have been made to investigate the nonlinear propagation of Alfv\'{e}n waves using numerical simulations.
\cite{mori04a} and \cite{mori04b} studied the dissipation process of Alfv\'{e}n waves in coronal loops. 
In their models, shear Alfv\'{e}n waves are driven by footpoint motion, and are coupled nonlinealy with fast and slow mode waves during propagation in the chromosphere and the corona.
They suggested that Alfv\'{e}n waves can create 1 million K corona when the r.m.s velocity of the footpoint motion is greater than 1 km s$^{-1}$.
For open flux tubes, \cite{kudo99} showed that Alfv\'{e}n wave heating model can also explain the spicule motion as well as it can transport sufficient energy flux to heat the corona. 
Moreover, \cite{suzu05,suzu06} showed that nonlinear low-frequency Alfv\'{e}n waves can simultaneously heat the corona and drive the fast solar wind.

One issue with the above simulations is the wave generation in the photosphere. 
Because there were only few observations on the time series of transverse motion at the photosphere \citep{tarb90}, the previous simulations
assumed an artificial velocity spectrum for wave generation.
As both the total power and the shape of the velocity spectrum can affect the wave generation and the heating process, it is necessary to
use observational data to justify whether Alfv\'{e}n waves are able to heat the corona.

It is possible to observe turbulent convection motion in the photosphere by Doppler analysis. Temporal velocity profiles for the vertical motion
have been studied for the sake of helioseismology \citep{harv93}.
\cite{carl97} used the observed Doppler velocity as a generation mechanism of
 acoustic waves and reproduced the \ion{Ca}{2} spectrum 
to compare with observations.
On the other hand, generation of Alfv\'{e}n waves requires transverse motion in the photosphere. 
Local correlation tracking (LCT) method \citep{nove88,titl89,berg98} reveals the horizontal motion of the granules.
Though many studies have been done on the spatial distribution of the horizontal velocity \citep{nove88,berg98}, few studies have been performed on the temporal velocity profiles \citep{tarb90}. 
This is because the LCT method is greatly affected by the atmospheric seeing.
However, a seeing free data set can be obtained by Solar Optical Telescope, SOT \citep{tsun08} on board Hinode \citep{kosu07},
so we can clearly understand the properties of the temporal profiles of the horizontal motion through LCT techniques.


An important physical process in the Alfv\'{e}n wave dynamics is the spicule dynamics.
One of the mechanism of spicule formation is slow mode shock model \citep{holl82a,shib82,suem82,ster88,ster00}. \cite{suem85} applied the slow shock model to fibrils and threads in a loop geometry. 
Leakage of p-mode from the photosphere eventually produce the slow mode shock
 \citep{suem90,depo04,hans06} . These shocks lift up the transition region resulting 
 in spicule motions. 
 However, spicule height cannot reach the observed value (5,000-7,000 km)
 only by the leakage of p-mode \citep{suem90}. \cite{kudo99} and \cite{sait01} considered 
 the slow mode generation through nonlinear Alfv\'{e}n wave mode conversion.
Thanks to \ion{Ca}{2} movies observed by Hinode satellite, detailed dynamics of the spicules have been understood \citep{depo07a,depo07,roup09}.
\cite{kim08} and \cite{he09} estimated the minimum wavelength of kink Alfv\'{e}n waves from horizontal motions of spicule. These observation can impose many restrictions on coronal heating models.

The purpose of the present study is to include a realistic horizontal velocity spectrum in the magnetohydrodynamic simulations 
and justify coronal heating by Alfv\'{e}n waves in the solar atmosphere.
In section 2, we will describe observation method to measure the horizontal photospheric motion.
The basic assumptions and our numerical model will be explained in section 3.
In section 4 we describe our numerical results and discussion will be included in section 5.
Finally, we will summarize our conclusion in section 6.

\section{Observations \& Data Reductions}

One of the main purposes of the Hinode satellite is to solve the coronal heating problem.
Among the three telescopes equipped on Hinode, Solar Optical Telescope (SOT) \citep{tsun08,suem08,ichi08,shim08}
is an optical telescope which has high spatial resolution (0.\arcsec2-0.\arcsec3) depending on wave length.
Since observations from space are not affected by atmospheric seeing, 
Hinode makes it easy to obtain a seeing free data set over a long time span.

\begin{figure}
\epsscale{.50}
\plotone{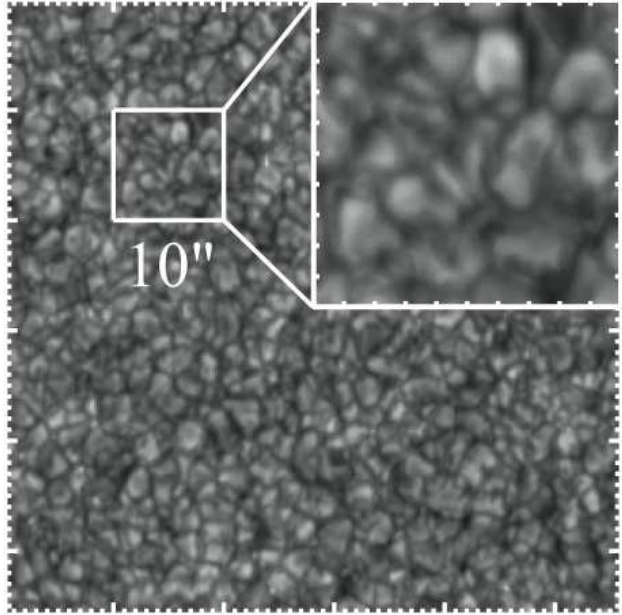}
\caption{
  Example of \textit{G}-band image observed by Hinode/SOT. 
  The image in the upper right square (10\arcsec$\times$10\arcsec) is zoom out image in the white square on the left.
 \label{fig1}}
\end{figure}

In this study, we used fourteen data sets from quiet regions observed with \textit{G}-band filter between 2006 October 31 and 2007 December 29.
We described the properties of each data set in Table \ref{tbl-1}.
The mean time cadence of each data set is less than 35 second. Basically, we selected the data sets whose duration is longer than 70 min.
Since every data set is selected to locate within $\pm$ 100 arcsec from the disc center, where the inclination angle is less than 6 degrees.
Pixel resolution of the CCD is 0.\arcsec054 and FOV is larger than 512 pixel $\times$ 512 pixel.
Figure \ref{fig1} shows the typical \textit{G}-band image used in our analysis. 

Dark current subtraction and flat-fielding were applied in the standard manner for all of the \textit{G}-band images. 
There are overall displacements between two images due to the solar rotation, tracking error, and satellite jitter.
We have calculated the cross correlation between two consecutive images and derived the displacement that had the highest correlation coefficient.
Using the displacement, we have adjusted all the data to an accuracy of sub pixel scale.

Generally, \textit{G}-band images also have intensity fluctuations from $p$ mode as well as granular motion.
In order to reduce the contribution from $p$ or $f$ modes, we have applied so called sub-sonic filtering technique \citep{titl89} to our data set.
The sub-sonic filtering technique removes the power where $\omega / k > V_{ph}$ in Fourier space so that the information moving faster than 
$V_{ph}$ does not appear in the movie.
In the present study, we have selected $V_{ph} = 5$ km s$^{-1}$.

\begin{figure}
\epsscale{.50}
\plotone{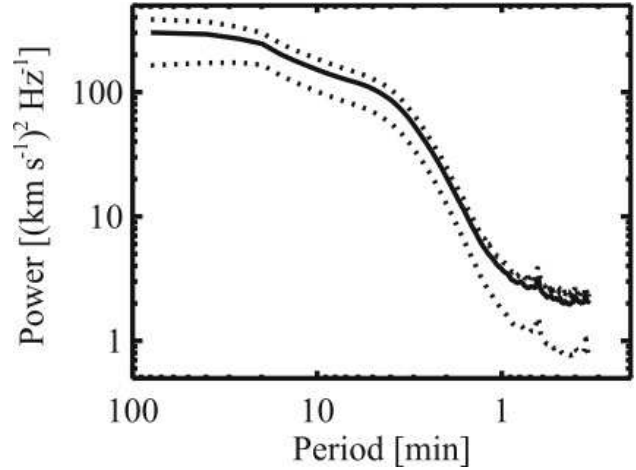}
\caption{
  Power spectrum density of the photospheric horizontal velocity.
  After the horizontal velocity is derived by the LCT technique, the spectrum is estimated by FFT.
  The solid line represents the mean power and dotted lines represent 1 sigma.
 \label{fig2}}
\end{figure}

In order to obtain the photospheric horizontal velocity, we have applied the LCT technique to the \textit{G}-band movies. 
The LCT techniques developed by \cite{berg98} uses rectangular subfields or "tiles" as a tracer and determines the displacement of tiles between two images. 
Since the horizontal velocity derived by the LCT technique usually depends on the size of the tracer, we have fixed the tile size to be 0.\arcsec 4.
Using the Fast Fourier Transform technique, we have estimated the one-sided power spectrum density, $P_{\nu}$ , from the time series of the photospheric horizontal velocity. 
The one-sided power spectrum density is defined as, 
	\begin{equation}
		<V_{\perp}^2> = \int _{\nu_{min}} ^{\nu_{max}}~P_{\nu}~d\nu \label{eq_power},
	\end{equation}
where $V_{\perp}$ is the horizontal velocity, and $\nu_{min}$, $\nu_{max}$ are the minimum and maximum frequencies constrained by the observational sampling time and duration. 
$<>$ means the temporal average. Power spectrum densities were first calculated at each fixed point, then we averaged over the FOV.
Figure \ref{fig2} shows an example of the power spectrum density in the region \# 1 listed in Table \ref{tbl-1}.
The solid line represents the mean power and the dotted lines above and below the solid line represent the range of 1 sigma.

We have also used \ion{Ca}{2} movie observed by Hinode/SOT on 22-Nov-2006 in order to compare apparent motion of spicules with our numerical simulation. The target region is the west limb over the quiet region. Pixel resolution of each image is 0.\arcsec054 and cadence is 8 sec.

\section{Numerical settings}
The basic concepts of our  simulations are the same as those of \cite{kudo99}.
We consider the nonlinear propagation of Alfv\'{e}n waves driven by photospheric convection.
The whole system is described by MHD equations with polytropic index $\gamma = $ 5/3, including gravity and empirical chromospheric cooling.
All physical variables are supposed to be the function of time $t$ and the distance $s$ along the flux tube measured from the photosphere.
Vector variables are composed of the component along the flux tube (poloidal component) and the toroidal direction.
We suppose the mean molecular weight to be 1, independent of the plasma status.
Moreover the poloidal configuration of the flux tube is assumed to be rigid, though the toroidal components of velocity ($V_{\phi}$) and magnetic field ($B_{\phi}$) are allowed to change with time.
Above assumptions lead the following basic equations.

	\begin{equation}
		{\partial \over \partial t} \left(  \rho A \right)+ {\partial  \over \partial s} 
		\left( \rho V_s A \right)= 0,
	\end{equation}

	\begin{eqnarray}
		{\partial \over \partial t} \left( \rho V_s A \right) + {\partial  \over \partial s} 
		\left( \left[ \rho {V_s}^2 + P_g + {B_\phi ^2 \over 8\pi} \right] A\right) \nonumber \\ = 
		\left( P_g + {{\rho V_{\phi}^2 \over 2}}\right) {dA \over ds} - \rho g_0 A {dz \over ds},
	\end{eqnarray}

	\begin{eqnarray}
		 {\partial \over \partial t} \left( \rho V_\phi A^{3/2} \right) + {\partial  \over \partial s} 
		\left( \left[ \rho V_\phi V_s - {B_\phi B_s \over 4\pi} \right] A^{3/2}\right) = 
		 A \rho L_{trq},
	\end{eqnarray}

	\begin{eqnarray}
		{\partial \over \partial t} \left( \sqrt{A} B_\phi \right) + {\partial  \over \partial s} 
		\left( \left[ B_\phi V_s - B_s V_\phi \right] \sqrt{A} \right)= 0,
	\end{eqnarray}

	\begin{eqnarray}
		&&{\partial \over \partial t} \left( \left[ {\rho |\mathbf{V}|^2 \over 2} + {P_g \over \gamma -1} + {|\mathbf{B}|^2 \over 8 \pi } \right] A \right) + \nonumber \\
		&&  {\partial  \over \partial s} 
		\left( \left[ \left\{ { \rho |\mathbf{V} |^2 \over 2} +{\gamma  P_g \over \gamma -1}+ {B_\phi ^2 \over 4\pi} 
		\right\} V_s - {B_\phi B_s V_\phi \over 4 \pi}\right] A \right) \nonumber \\ &&= 
		-L_{rad} A - {\rho V_s g {dz\over ds}} A + \rho V_{\phi} \sqrt{A} L_{trq},
	\end{eqnarray}

	\begin{equation}
		P = {\rho k_B T \over m_p},
	\end{equation}

where  $r$ is the radius of the flux tube, $A$ is the cross section of the flux tube,
 $\rho$ is the mass density, $P_g$ is the gas pressure,
 $T$  is the temperature, $V_s$ is the poloidal component of the velocity, $V_\phi$ is the toroidal 
 component of the velocity, $L_{trq}$ is the photospheric torque used as a wave driver, $B_s$ is the poloidal component of the magnetic field, and $B_{\phi}$
  is the toroidal component of the magnetic field. 
We include the chromospheric cooling $L_{rad}$ according to \cite{hori97} only when the plasma temperature is less than $8\times 10^5$ K. This method involves two processes.
For plasmas whose temperature is less than $4 \times 10^4$ K, we implement the empirical cooling for optically thick plasmas \citep{ande89,ster93}, 

	\begin{equation}
		L_{rad} = 4.9 \times 10^9 \rho ( s,t),~~\mathrm{erg~cm^{-3}~K^{-1}} .
	\end{equation}
For plasmas whose temperature is more than $4\times10^4$ K and less than $8\times 10^5$ K, we use the cooling for optically thin plasmas \citep{hidn74,ster93}

\begin{figure}
\epsscale{.50}
\plotone{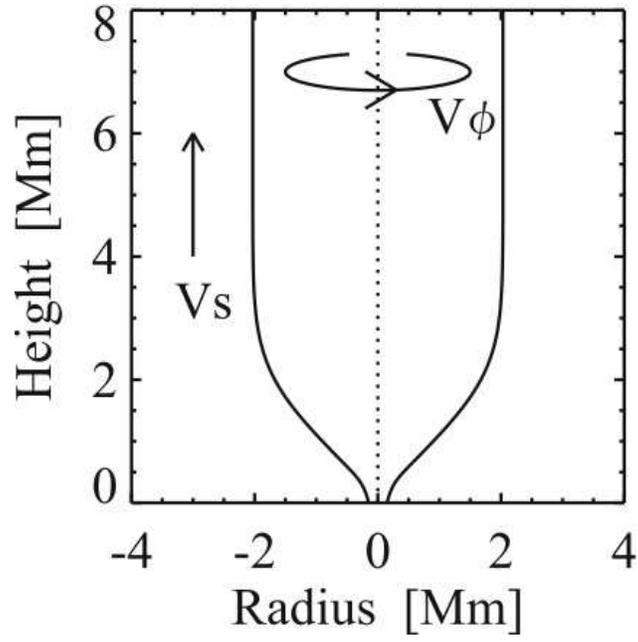}
\caption{
	Structure of the flux tube that we assumed in our model.
 \label{fig3}}
\end{figure}

Although flux concentrations at the photosphere are on the order of the pressure scale height, they rapidly expand in the upper atmosphere due to exponential decrease of gas pressure.
If we assume mechanical equilibrium between the magnetic pressure of the flux tube and the ambient gas pressure, plasma beta is nearly 1 in the photosphere and the chromosphere.
In order to reproduce such a situation, we assume the same structure of the flux tube as that of \cite{kudo99},

	\begin{equation}
		r = \int \cos \alpha ds,
	\end{equation}
	\begin{equation}
		z = \int \sin \alpha ds,
	\end{equation}
where
	\begin{equation}
		\alpha = \alpha _t + (\alpha _r - \alpha _t ) f_n,
	\end{equation}
	\begin{equation}
		\alpha _r=  -\arctan \left( {-4 H_0 \over r} \right),
	\end{equation}
	\begin{equation}
		\alpha _t=  -\arctan \left[ { (z/z_d)}^2 \right],
	\end{equation}
	\begin{equation}
		f_n=  -{1\over 2} \{ \tanh [  (z-0.2z_d)/(0.1z_d)] -1\}.
	\end{equation}
$H_0$ is pressure scale height of the photosphere (150 km), $z_d$ is equal to 12 $H_0$ (1800 km) , and $r$ is the radius of the flux tube ($A \propto r^2$).
The poloidal field $B_s$ is determined by $B_s = B_0 (r_0/r)^2$, where $r_0$ is the radius of the flux tube at the photosphere ($r_0=H_0=150$ km), and
$B_0$ is defined so that plasma beta equals to 1 in the photosphere (Fig. \ref{fig3}).

\begin{figure}
\epsscale{0.5}
\plotone{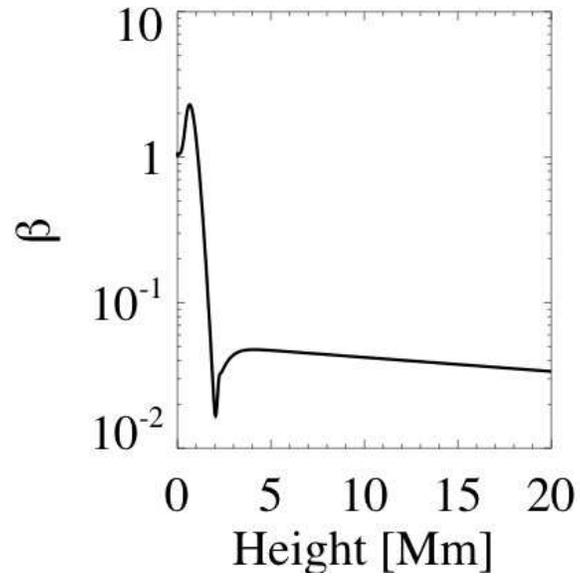}
\caption{
	Profile of plasma beta with respect to height.
 \label{fig4}}
\end{figure}

The initial atmosphere is supposed to be magnetohydrostatic equilibrium with constant gravity ($g = 2.7\times 10^4$ cm s$^{-2}$). 
The temperature is taken to be as follows. 

	\begin{equation}
		{T\over T_0} = 1 + {1\over 2} (a_c-1) \left( 1+ \tanh{{z-15 H_0 \over0.5 H_0}} \right),
	\end{equation}
where $a_c = 300$, and $T_0$ is the photospheric temperature (5000 K).
The density at the photosphere is supposed to be $2.53 \times 10^{-7}$ g cm$^{-3}$.
Figure \ref{fig4} shows the profile of the plasma beta with respect to height in our model atmosphere.

The foot point of the flux tube is subjected to an external torque ($L_{trq}$) 
so as to drive the Alfv\'{e}n waves.
The formulation of the external torque is

	\begin{equation}
		L_{trq} = {1\over 2} \times F_0 \times \left( 1 - \tanh{{z-0.75H_0 \over 0.075H_0}} \right).
	\end{equation}
The amplitude of the torque, $F_0$,  is determined so that the resulting toroidal velocity of the foot point becomes the given velocity time series that reproduces the given velocity spectrum.

The numerical domain ranges from the photosphere to more than 900 $H_0$ (= 1.35 $\times 10^5$ km).
The grid size below 110 $H_0$ from the photosphere is 15 km while the grid size over 110 $H_0$ becomes 1.01 times larger per grid. Since wave amplitude increases exponentially with height in the stratified atmosphere, we set the gravity to be 0 above 900 $H_0$
in order to reduce the effect of reflection at the upper boundary.

Our numerical scheme is an HLLD approximate Riemann solver developed by \cite{miyo05}.
Since the HLLD scheme has positivity in some situations, this scheme is more robust than a Roe-type linear Riemann solver.
Moreover this HLLD scheme has good resolution and numerical efficiency.
The spatial accuracy obtains 2nd order by using the TVD-MUSCL method while the temporal accuracy obtains 2nd order by using the Runge-Kutta method.
We expect that a conservative scheme like the HLLD scheme can capture the shocks better than the non-conservative scheme used in \cite{kudo99}.
Conservative schemes are suitable for the situation when Alfv\'{e}n waves are converted into compressional waves resulting in a lot of shock formation.

\section{Results}
We have investigated the relation between the initial Alfv\'{e}n wave spectrum and the nonlinear behavior of the upper atmosphere.
After assuming the model atmosphere, Alfv\'{e}n waves are generated by torque at the foot point of a magnetic flux tube.
First of all, monochromatic waves that have 1 km s$^{-1}$ amplitude at the photosphere are generated.
Then, we have driven Alfv\'{e}n waves that have continuum spectrum with different wave amplitudes.
Finally, the observed velocity spectrum is included in our numerical simulation as a wave driver.

\begin{figure}[p]
\epsscale{1.0}
\plotone{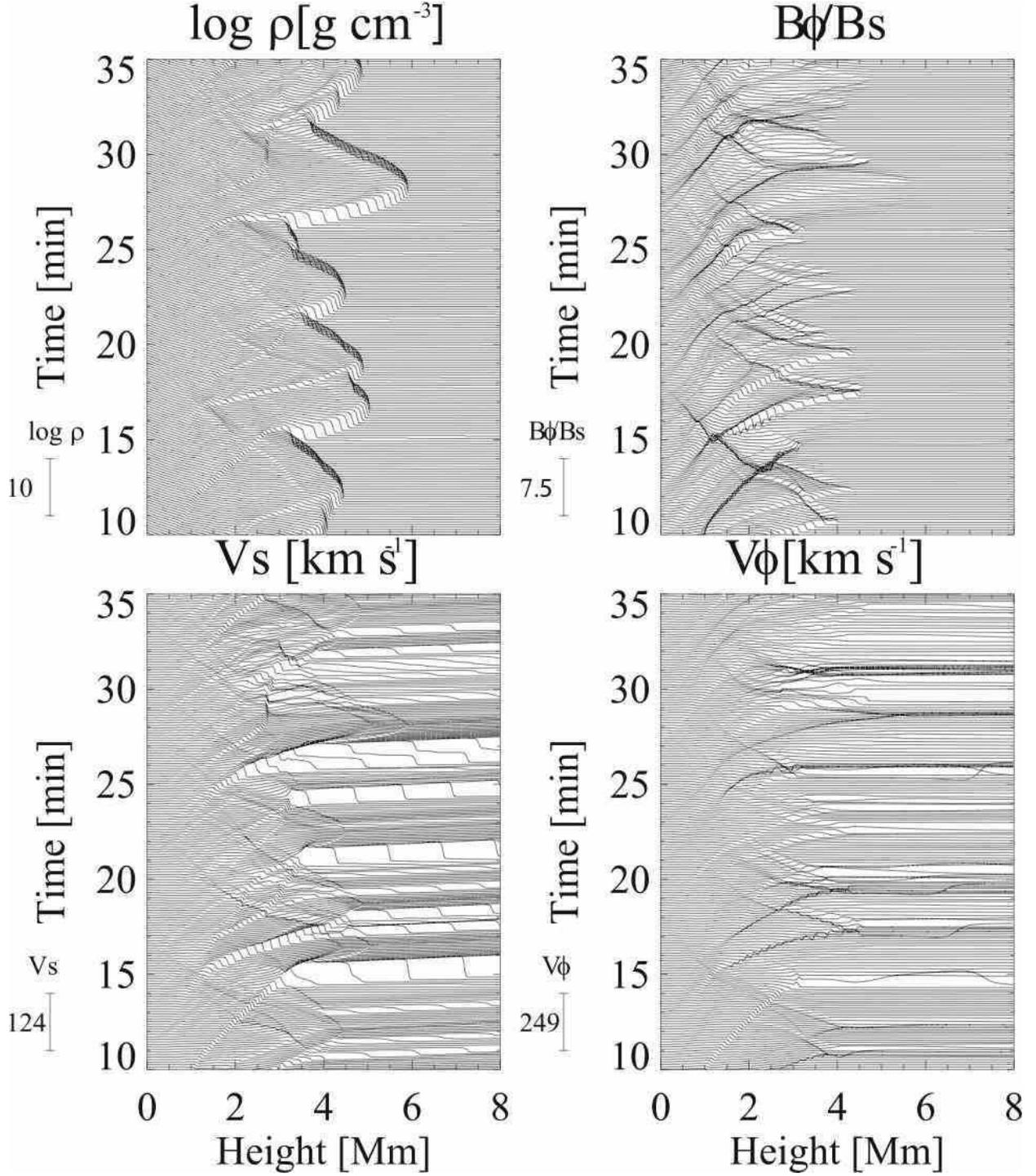}
\caption{
	Dynamics of the Alfv\'{e}n waves in our typical simulation. 
	Plots in each panel are stacked with time increasing upward 
	in uniform increments of 8.0 s.
	Each panels represents the time profile of density (upper left), 
	B$_{\phi}$/Bs (upper right), V$_s$ (lower left), and V$_{\phi}$
	(lower right).
 \label{fig5}}
\end{figure}

Figure \ref{fig5} shows the dynamics of Alfv\'{e}n waves in typical case of our numerical
 simulation.
Plots in each panel are stacked with time increasing upward in uniform increments of 8.0 s.
Although we only generated toroidal motion, poloidal motion can be seen in the figure.
This is the result of mode conversion from Alfv\'{e}n waves to slow mode waves.
The slow mode waves push up the transition region, which can be observed as spicules \citep{kudo99}.

\subsection{Monochromatic wave}
We generated monochromatic Alfv\'{e}n waves, changing the wave period from 10 sec to 750 sec.
For each period, the amplitude is fixed to be 1 km s$^{-1}$ and we measured the resulting energy flux in the corona (15,000 km) and 
the height of the transition region.

\begin{figure}
\epsscale{.50}
\plotone{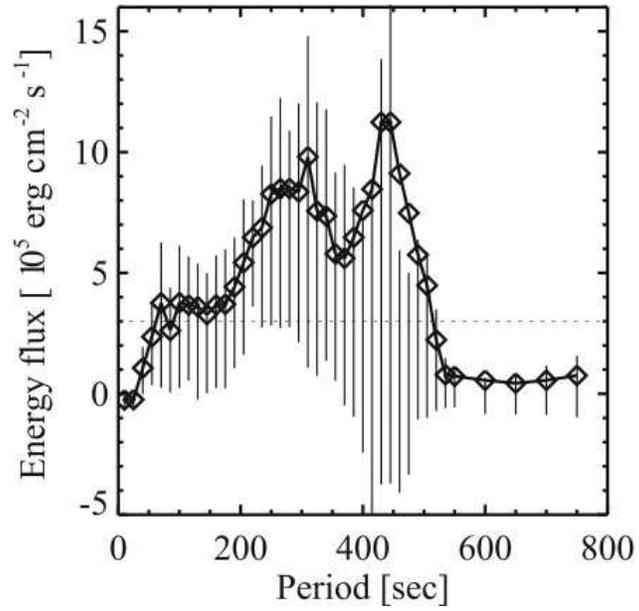}
\caption{
  Total energy flux measured at the corona with respect to the wave period.
  The amplitude of each wave period is fixed to 1 km s$^{-1}$.
  The white diamonds show each data point respectively while the vertical solid lines at each diamond represent 
  the range of 1 sigma. 
  The horizontal dotted line indicates the sufficient energy to maintain the corona ($3\times10^5$ erg cm$^{-2}$ s$^{-1}$). 
 \label{fig6}}
\end{figure}

Figure \ref{fig6} shows the relationship between the mean total energy flux measured at the corona and the wave period.
Total energy flux is described as follows;
	\begin{equation}
		F=V_s \left[ {1\over 2} \rho ( V_s^2+V_{\phi}^2) + {\gamma \over \gamma -1} P_g + \rho g z + {B_{\phi}^2 \over 4 \pi} \right] - {B_{\phi} B_s \over 4 \pi} V_{\phi}.
	\end{equation}
The dotted line represents the sufficient energy flux to maintain the hot corona , $3\times10^5$ erg cm$^{-2}$ s$^{-1}$ \citep{with77}.
The vertical solid lines at each data point represent the error bar for 1 sigma.
The energy flux starts to decrease rapidly when the wave period becomes larger than 500 sec.
The time span, 500 sec,  roughly corresponds to the Alfv\'{e}n wave travel time back and forth between the photosphere and the transition region.
Therefore the waves whose period are longer than 500 sec are canceled out when they come back to the photosphere.
The waves whose period are shorter than 60 sec also start to decrease their energy flux at the coronal height.
In the case of the shorter period Alfv\'{e}n waves, the shorter period longitudinal waves are frequently produced through nonlinear mode conversion.
Because such waves are large in number and have shorter damping length, a large amount of wave power is dissipated in the chromosphere.
Wave periods that efficiently heat the corona range from 60 to 500 sec.
One of the reason for that is the Alfv\'{e}n wave resonance between the photosphere and the chromosphere, which we discuss in detail in the following section.
The energy flux profile with respect to wave period (Fig \ref{fig6}) clearly shows resonant peaks at 300 and 450 sec.

\begin{figure}
\epsscale{.50}
\plotone{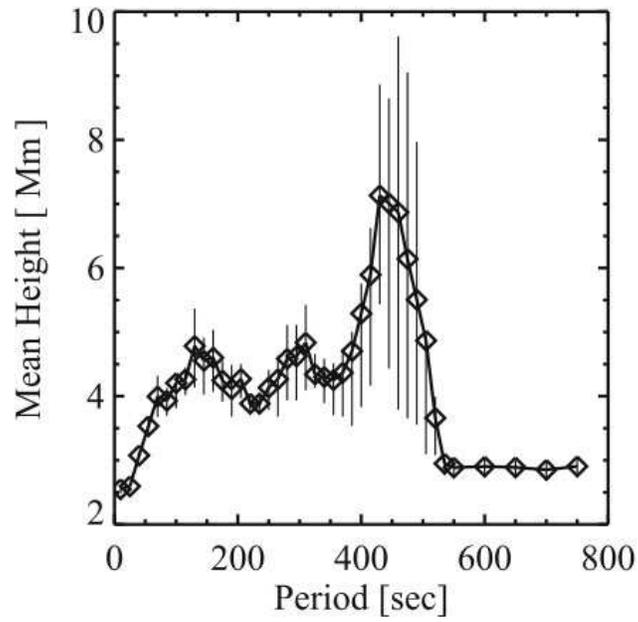}
\caption{
  Transition region height with respect to the wave period.
  The amplitude of each wave period is fixed to 1 km s$^{-1}$.
  The white diamonds show each data point respectively while the vertical solid lines at each diamond represent 
  the range of 1 sigma. 
 \label{fig7}}
\end{figure}

Figure \ref{fig7} shows the relationship between the mean transition region height and the wave period.
In this study the transition region is defined as the region of the atmosphere at a temperature of $8\times 10^5$ K.
Similar to the mean energy flux, the waves whose period is shorter than 60 sec or longer than 500 sec cannot create the tall spicules that are frequently observed in the real Sun.
The transition region height profiles with respect to wave period (Fig. \ref{fig7}) also represent the resonant peaks at the 130, 300, 450 sec.

\subsection{The waves that have continuum spectrum}
We have investigated the dependence of nonlinear Alfv\'{e}n wave propagation on the spectral shape of the initial waves.
Two different wave spectrums, white noise and pink noise, are considered as wave generators.
The power spectrum densities $P_{\nu}$ defined by equation (\ref{eq_power}) of the white noise and pink noise
are proportional to $\nu ^{0}$ and $\nu ^{-1}$ respectively.
The frequency range is limited from $6 \times 10^{-4}$ to $5 \times 10^{-2}$ Hz in the same way as that of \cite{suzu06}.
For each spectrum, the amplitude of the waves changes between 0.1 and 5 km s$^{-1}$.

\begin{figure}
\epsscale{.50}
\plotone{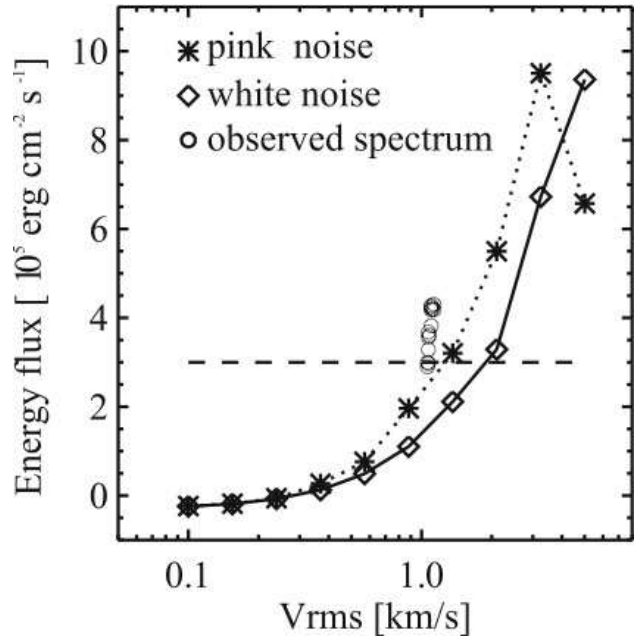}
\caption{
  Total energy flux measured at the corona with respect to the wave amplitude.
  The solid line with white diamonds shows the white noise case 
  while the dotted line with asterisks shows the pink noise case.
  The white circle shows the observed spectrum case.
  The horizontal dashed line indicates the sufficient energy to maintain the corona ($3\times10^5$ erg cm$^{-2}$ s$^{-1}$). 
 \label{fig8}}
\end{figure}

Figure \ref{fig8} shows the dependence of the total energy flux in the corona on the amplitude of the wave.
The horizontal dashed line represents the sufficient energy flux to heat the corona.
The diamond and asterisk symbols show the case for white noise and pink noise respectively.
Generally energy flux at the corona increases with the wave amplitude.
In white noise case, we need a wave amplitude more than 2 km s$^{-1}$ to maintain the hot corona while in the pink noise case, the necessary wave amplitude is more than 1.3 km s$^{-1}$.
Pink noise waves heat the corona more efficiently than white noise waves because the pink noise waves include more power in the lower periods around 300 sec.
Around these lower period ranges, Alfv\'{e}n wave resonant periods exist.
Therefore large amount of energy flux can propagate to the corona without reflection.
In the case of pink noise, energy flux above the 5 km s$^{-1}$ wave amplitude decreases because a large fraction of the wave energy is dissipated in the chromosphere.
When the wave amplitude becomes larger, the nonlinearity of the waves also becomes also larger.
Then the waves experience frequent nonlinear mode coupling and dissipates rapidly.

\begin{figure}
\epsscale{.50}
\plotone{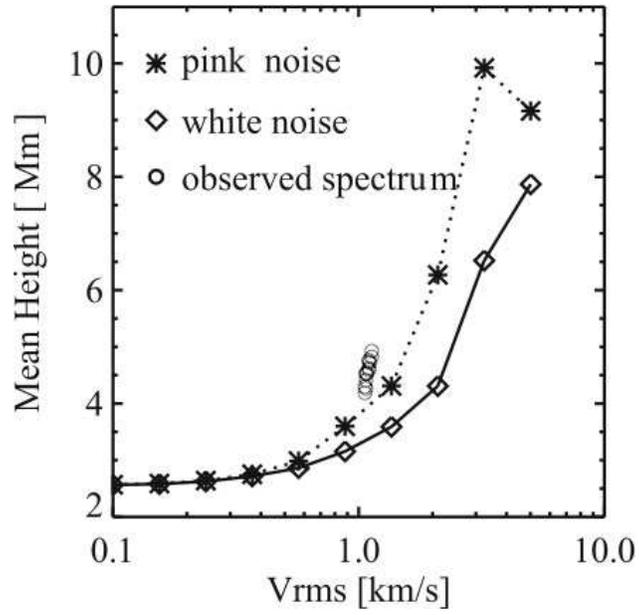}
\caption{
  Transition region height with respect to the wave amplitude.
  The solid line with white diamonds shows the white noise case 
  while the dotted line with asterisks shows the pink noise case.
  The white circle shows the observed spectrum case.
 \label{fig9}}
\end{figure}

Figure \ref{fig9} reveals the relationship between the transition region height and the amplitude of the waves.
For the same reason described above, the spicule height is taller in the pink noise case than in the white noise case.

\subsection{The waves that have the observed spectrum}
We have derived the photospheric horizontal velocity from the \textit{G}-band movies by using LCT techniques.
According to the figure \ref{fig2}, the observed velocity spectrum generally has a  double power law distribution.
For the power spectrum density derived from the different regions listed in the Table \ref{tbl-1}, we have performed a the double power law fitting.
The fitting function is described as follows.
	\begin{eqnarray}
		P_{\nu} \propto
		\left\{
			\begin{array}{c}
				 \nu ^{\alpha _L}    ~~~(\nu < \nu _b)    \\
				\nu ^{\alpha _H}   ~~~(\nu \ge \nu_b).
			\end{array}
			\right.
	\end{eqnarray}
The four parameters that define the spectral shape ( square root of the total power, $\sqrt{<V_{\phi}^2>}$, spectral break frequency, $\nu _b$, 
and the two power index, $\alpha _L, \alpha _H$) are described in the Table \ref{tbl-1}.
In order to distinguish the high and low frequency components, we implement the Fourier filter with a spectral break frequency.
As a result, the low frequency components correspond to the granular motion while the high frequency components correspond to the inter granular motion. Since the power spectrum density in high frequency part is much smaller than that of low frequency part, waves driven by the granular motion is likely to be important for coronal heating.

\begin{figure}
\epsscale{.50}
\plotone{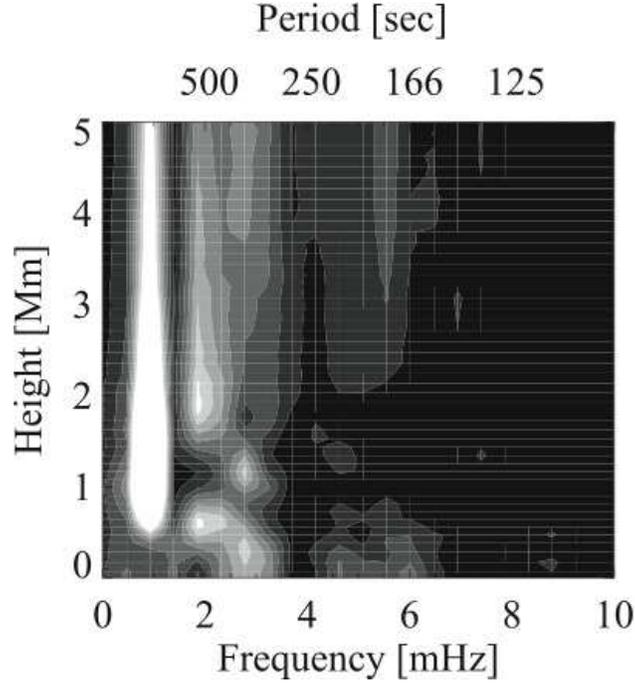}
\caption{
  Power spectrum density of toroidal velocity measured at each height.
  The power spectrum is normalized at each height so that the total power becomes one.
  The initial wave spectrum corresponds to the observed spectrum \#1 listed in Table \ref{tbl-1}.
 \label{fig10}}
\end{figure}

Then we have included the observed velocity spectrum listed in the first line of the Table \ref{tbl-1} into our numerical simulations.
In this case, the total energy flux becomes $4.2 \times 10^5$ erg cm$^{-2}$ s$^{-1}$ and the mean spicule height becomes 4.7 Mm. The results for the other case in the Table \ref{tbl-1} are plotted 
in the figure \ref{fig8} and \ref{fig9} as white circles.
Figure \ref{fig10} is the Fourier spectra of the toroidal velocity with height.
The color represents the power normalized by the total power at each height.
There are at least three resonant peak at 1,3, and 5 mHz.

\subsection{Synthesized spicules}
\begin{figure}
\epsscale{1.0}
\plotone{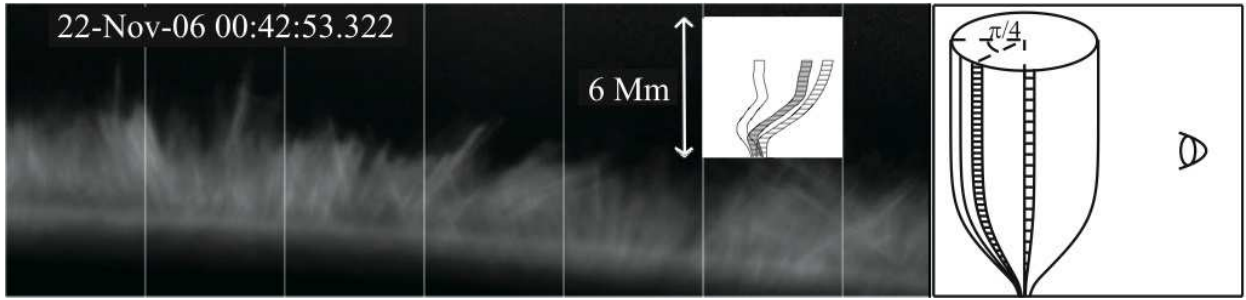}
\caption{
	Synthesized spicules compared with \ion{Ca}{2} spicules observed by Hinode/SOT.
	Snapshot image from our simulation is shown in the upper right white square 
	in the left panel.
	The distance between vertical white lines is 6 Mm.	
	The right panel shows the initial positions of three synthesized spicules with respect 
	to an observer. Three bands in the panel represent synthesized spicules that 
	have different angle against the line of sight. 
 \label{fig11}}
\end{figure}

Although our numerical model is just 1 dimensional model along the flux tube, this flux tube has a three dimensional structure in real 3 dimensional space.
Therefore we can synthesize the spicule motion in 3 dimensional space if we trace the field lines which have cool (less than $1\times 10^5 $ K) plasma.
In order to include projection effects, synthesized spicules are plotted in three different orientations with respect to an observer (see the right panel of figure \ref{fig11}).
Although three bands of the spicules in figure \ref{fig11} are synthesized from the same simulation, these spicules are viewed from different angles.
The spicule with the white band is initially set on the plane that is made by the vertical direction and the line of sight direction, while the spicule with the thin shaded band is initially set on the sky plane.
The spicule with the thick shaded band is initially set on the plane which is rotated 45 degree around vertical axis from the sky plane.

We have compared the observed  \ion{Ca}{2} spicules with the synthesized spicules in the left panel of figure \ref{fig11}. We superimposed the snapshot of the synthesized spicules on the \ion{Ca}{2} image. 
Figure \ref{fig11} is also available as an mpeg animation in the electronic edition of the \textit{Astrophysical Journal}. 
From the movie, we can easily compare the height and the structure of spicules. The height of spicules from the photosphere deduced from the \ion{Ca}{2} movie by eyes are comparable with that of synthesized ones. On the other hands, the movie clearly shows that the synthesized spicules are bent due to Alfv\'{e}n wave propagation, although the observed spicules usually have very straight structures. In the movie, white, thin shaded, and thick shaded spicules are changed into blue, black and red bands respectively. Although our model does not essentially depend on the foot point radius of the flux tube, appearance of the spicules can change.
So we include the three different cases synthesized from three different foot point radius, 
$r_0 = 15, 75,150$ km in the online movie.

\section{Discussion}
\subsection{Alfv\'{e}n wave resonant cavity}
Alfv\'{e}n waves are driven by the magneto-convection at the photosphere.
The previous studies \citep{kudo99,mori04a,mori04b,suzu05,suzu06} have assumed artificial wave drivers such as white noise spectrum
because there were few observations of the photospheric temporal spectrum.
In the present study, we include the observed disturbance into our numerical simulations.

To investigate the dependence of coronal heating on the wave period, we have implemented a parameter survey on the wave period.
As a result, we have found that the waves of period around 100-500 sec can transport the large amount of wave energy
to the corona.
These special period waves are also found to make the height of the transition region taller. 
Since the observed spectrum has large power around 300 sec, it is more efficient for coronal heating and spicule formation over 
white noise spectrum.
\cite{anto08} also reported that the coronal temperature becomes highest when they generate the monochromatic Alfv\'{e}n waves of the period of 300 sec.

The reason for the response of our model atmosphere against the special wave periods is considered to be Alfv\'{e}n wave resonance 
between the photosphere and the transition region.
Linear analysis by \cite{holl81} and \cite{lero81} suggested that the Alfv\'{e}n wave resonance can occur even when the flux tube is not closed but opened.
In the present study, our bottom boundary works as a reflective boundary since we have imposed the rigid boundary at the photosphere.
The transition region also becomes a reflective boundary because the Alfv\'{e}n speed increases rapidly.
Therefore standing waves appear in the region between the photosphere and the transition region and the standing waves will acquire 
larger amplitude by Alfv\'{e}n wave resonance.

\begin{figure}
\epsscale{0.5}
\plotone{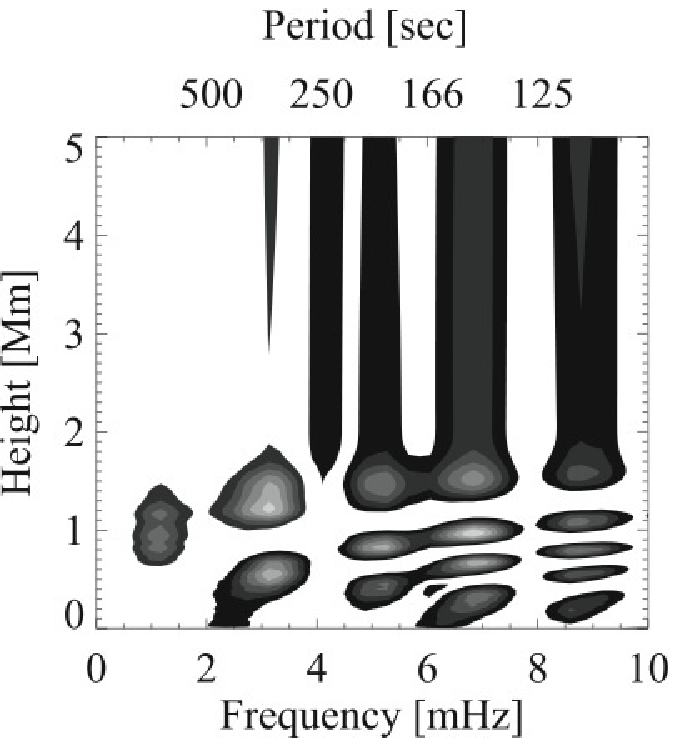}
\caption{
  Power spectrum density of toroidal velocity measured at each height.
  The power spectrum is normalized at each height so that the total power becomes one.
  The Alfv\'{e}n wave is driven by white noise with 0.1 km s$^{-1}$ amplitude.
 \label{fig12}}
\end{figure}

Let us examine the linear properties of Alfv\'{e}n resonance in a flux tube in our model atmosphere by assuming a small wave amplitude using the same 1.5D MHD simulations. We assume small amplitude waves with white noise spectrum to check the Alfv\'{e}n wave resonance. 
Figure \ref{fig12} is Fourier spectrum of the toroidal velocity and it clearly shows resonant behavior.
The color represents the normalized power of the toroidal velocity at each height.
The fundamental mode appeared around 1 mHz while the second harmonics appeared around 3 mHz.
There are a lot of higher harmonics below 300 sec.
Since the height of the transition region changes according to the wave property, resonant frequency will also change 
depending on the situation in the real Sun.
When we include the observed velocity spectrum, the fundamental mode, $\sim$ 1000 sec, has biggest power through the whole atmosphere.
If the Alfv\'{e}n wave resonant cavity actually exists, the resonant periods are expected to be observed as spicule motion or coronal transverse velocity.

In our simulation, the photospheric boundary is a fixed boundary by definition 
while the transition region works as a free boundary because of the high Alfv\'{e}n 
velocity at the corona. This means that the resonance occurs between the node (photosphere) 
and antinode (transition region). If, for simplicity, we assume the constant Alfv\'{e}n velocity ($V_A$) 
in the resonance region
, the resonance frequency $\omega _n$ can be 
described as follows,
	\begin{equation}
		\omega _n = {V_A \over 4 L } \left( 2n-1 \right),~(n=1,2,3...) ,
	\end{equation}
where L is the transition region height. When L equals to 2,000 km, and $V_A$ equals to 8 km s$^{-1}$, resonance frequency becomes, 1,3,5 ... mHz. These frequencies agree quite well with 
our numerical experiments (Figure \ref{fig12}).

In the present study, we assume a reflective boundary at the photosphere.
This assumption may not be valid if the flux tube extends straight down below the photosphere.
However in the sub-photosphere, Alfv\'{e}n speed may be smaller than in the upper atmosphere and the flux tube can be easily bent by convection motion.
In that case, the wave reflection can eventually occur at some point along the flux tube due to inhomogeneity of phase speed, 
resulting in a change in the resonant frequency.
In order to confirm the boundary effect, we performed another simulation including 
the subphotospheric layer where 
the Alfv\'{e}n velocity decreases exponentially with depth. It was found that even if the additional 
subphotospheric layer exists, Alfv\'{e}n wave dynamics and energetics do not change qualitatively or quantitatively.

\subsection{Photospheric horizontal velocity spectrum}
We have investigated the granular motion by using LCT techniques for 14 different regions.
LCT techniques require a stable data set with no effects from atmospheric seeing.
The spatial resolution of Hinode/SOT is the highest of space optical observation of the Sun.
For example, Solar Optical Universal Polarimeter (SOUP) aboard Spacelab 2 was a telescope with 30 cm aperture \citep{titl86} 
while Hinode/SOT is a telescope with 50 cm aperture.
Although some ground-based observations have higher spatial resolution than that of Hinode/SOT, there are few data sets of this quality
even if adaptive optics are used. 

In general, the results obtained in the present study are almost the same as that of \cite{tarb90}.
We want to note that the derived LCT velocity is the apparent velocity, even though we have carefully removed the effect of $p$-mode.
\textit{G}-band intensity contrasts are considered to change due to the destruction of CH molecules when the plasma density decreases 
or the temperature increases \citep{stei01}. It is possible that the 
LCT velocity includes these intensity contrasts as well as the granular motion.
Since recent realistic 3D simulation \citep{stei98,carl04} succeeded in reproducing granular motion, detailed comparison of LCT velocity 
with the simulated velocity should be investigated in the future.
In order to avoid the ambiguity of LCT velocity, it is possible to calculate velocity by feature tracking \citep{berg98,mull94} but we do not consider this method.

Another possibilities to generate the Alfv\'{e}n waves are rapid changes in the magnetic topology caused by reconnection or interaction between 
flux tubes and acoustic waves coming from ambient non-magnetic atmosphere \citep{hasa05,hasa08}.
Recent observations reveals that granular-sized transient horizontal fields are emerging in quiet region \citep{ishi08}.
\cite{isob08} found from 2D and 3D MHD simulation that the high frequency Alfv\'{e}n waves can be created by magnetic reconnection between these transient fields and
ambient vertical fields.

Only the toroidal velocity is included in our simulations whereas the observed velocity is lateral.
This may suggest that we should include the rotation of the observed velocity rather than lateral one.
However, the basic equation of lateral motion is the same as that of toroidal motion except for the centrifugal force term.
Since the centrifugal force turned out to be negligibly small in our simulations, it is valid to use troidal motion as a wave driver.

\subsection{1.5 D approximation}
We have assumed a 1 dimensional model along the flux tube from the photosphere to the corona. Generally, there should be leakage of Alfv\'{e}n wave energy in the flux tube 
through the conversion to fast mode waves.
Moreover, in the real Sun, there appears to be other multi-dimensional effects such as Alfv\'{e}n \& fast mode coupling by field bending \citep{went74}, 
phase mixing \citep{heyv83}, and Alfv\'{e}n wave resonance \citep{ions78}.
The synthesized spicules have bending structures due to Alfv\'{e}n wave whereas the observed spicules usually looks very straight. 
Indeed, wave lengths of kink Alfv\'{e}n waves propagating along spicules are estimated to be more than 45,000 km \citep{kim08} or 
2,000-4,000 km \citep{he09}.
Since the bending structures in our simulation are usually $\sim$ 1,000 km, there should be another dissipation mechanism of
Alfv\'{e}n waves in multi-dimensional space.

One possibility for multi-dimensional dissipation that can be suggested by our simulations is instabilities due to the standing waves
\citep{heyv83} in the Alfv\'{e}n wave resonant cavity.
Shear flows at the antinodes of the standing waves can drive the Kelvin-Helmholtz instability, causing them to decay in a few periods.
Moreover, current density at the nodes of the standing waves can cause the tearing mode instability.
How much energy can be dissipated in such mechanisms will be interesting topics for future study.

If there is greater Alfv\'{e}n waves dissipation in the chromosphere than the present model, 
the spicule heights must be reduced to some extent due to decrease in nonlinearity.
Although pure $p$-mode injection cannot explain the realistic spicule height \citep{suem90},
$p$-mode may be candidate for additional momentum in spicules.
Slow modes can be created by reconnection \citep{take01} or convective collapse \citep{steiner98,take99}.
\cite{naga08} observed the transient down flows with intensification of the field strength predicted by the model of convective collapse.
These down flows are expected to be bounced back to the upper atmosphere, resulting in a wave source for slow mode \citep{ster88,ster89}.

Recently, various simulations for nonlinear MHD wave propagation in multi-dimensional
 space have been done. \cite{shib83} studied the nonlinear propagation of magnetoacoustic waves in the solar chromosphere in 1D/2D space and 
 found that 1D approximation is reasonable when the plasma beta is lower than unity.
The propagation of nonlinear Alfv\'{e}n waves were studied as a 
mechanism for solar and astrophysical jets \citep{shib85}.
For coronal heating problem, \cite{hasa05} and \cite{hasa08} implemented the 2-D simulations of magnetoacoustic waves propagating in a flux tube.
Moreover, \cite{stei98} found the bending and horizontal displacement of a flux sheet 
through the coupling with the convective motion self-consistently with radiative MHD code.
These calculations only include the region below the chromosphere because the dynamical time scale becomes very short in the corona.
Although recent 3D MHD simulations including the corona with 
fine spatial resolution can sufficiently resolve the wave dissipation and 
include the physics such as thermal conduction or radiative transfer \citep{gudi05,leen09}, it is 
worthwhile investigating 1D simulation to understand nonlinear wave propagation.

\subsection{Energy equation}
When we want to precisely investigate the energy transfer between the photosphere and the corona, it is necessary to include 
radiative transfer and thermal conduction.
The cooling time in the photosphere and the chromosphere changes from 1 - 1,000 sec depending on height \citep{stix70} while 
we assume a constant cooling time in the present study.
Therefore the distribution of the temperature could be changed if we properly include radiative loss.
Though the compressional mode waves are affected by radiative damping, it is inferred that non-compressional modes like Alfv\'{e}n waves 
are not significantly damped.
Moreover, since the cooling time in the chromosphere ($\sim$ 200 sec) is much longer than the Alfv\'{e}n crossing time ($\sim$ 20 sec), 
the chromospheric dynamics of our simulation would not be greatly affected if radiative transfer was included.

Thermal conduction is not so important as long as we are interested in the spicule dynamics. In our model, spicules are created by collision between transition region and slow mode shocks. The slow mode shocks are generated by nonlinear mode conversion from Alfv\'{e}n waves in the chromosphere. These formation process of spicules must not be affected by thermal conduction because the chromosphere is too cool. Thermal conduction become important when we want to study the energy balance in the corona or the transition region.

\cite{suzu05,suzu06} performed similar simulations to ours but including thermal conduction and radiative cooling.
Their model succeeded in maintaining the corona and driving the fast solar wind by nonlinear low-frequency Alfv\'{e}n wave.
Since the observed photospheric velocity spectrum is similar in shape and total power to the wave spectrum assumed in \cite{suzu06},
their model can maintain the corona when including the observed velocity spectrum.
However the plasma beta in the photosphere and the chromosphere is very large in their simulation, therefore the dynamics, such as spicule motion, must be different.
It is necessary to assume more realistic magnetic field distribution along the flux tube to explain spicule formation and coronal heating simultaneously.

\cite{carl97} derived the \ion{Ca}{2} spectrum by using hydrodynamic simulations with non-LTE radiative transfer.
In order to compare simulated results with observations, it is important to synthesize the spectrum during or after the simulations.
Numerical simulations with radiative transfer are also needed when we solve the convection layer.
Magnetic intensification through convective collapse and flux tube buffeting by convection motion 
can be included self-consistently by the radiative MHD simulations \citep{steiner98}.

\section{Conclusion}
We have investigated the nonlinear evolution of Alfv\'{e}n wave propagation along a flux tube in the solar atmosphere. 
In particular, we have concentrated on the dependence of the energy flux transmission and the spicule formation on the 
Alfv\'{e}n wave period.
Monochromatic Alfv\'{e}n waves with given amplitude or Alfv\'{e}n waves that have continuum spectrum with various amplitudes are
generated in our model atmosphere.
Moreover we have analyzed the photospheric horizontal velocity spectrum from \textit{G}-band movies and included the spectrum 
in our numerical simulations.
The results obtained in this paper are summarized as follows;

\begin{enumerate}
\item 
The region between the photosphere and the transition region can be an Alfv\'{e}n wave resonant cavity.
When including the observed velocity spectrum, we have obtained at least three resonant 
frequencies, 1,3, and 5 mHz.
The Alfv\'{e}n waves that have resonant frequencies transfer more energy to the corona and increase the spicule height.

\item 
Pink noise ($P_{\nu} \propto \nu ^{-1}$) has an advantage for coronal heating over white noise ($P_{\nu} \propto \nu ^{0}$).
Alfv\'{e}n waves generated by pink noise also increase the spicule height.

\item
The photospheric horizontal velocity spectrum has larger power around 3 min than pink or white noise.
The power around 3 min or more comes from granular motion while the high frequency components come
from inter-granular motion.

\item
The observed velocity spectrum has an advantage for coronal heating over white noise.
Obtained energy flux is $4.2\times 10^5$ erg cm$^{-2}$ s$^{-1}$, which means that Alfv\'{e}n waves transfer sufficient energy to heat the quiet corona. 

\item
Structures of spicules synthesized from our numerical simulation are bending more than observed \ion{Ca}{2} spicules.
This implies the dissipation of Alfv\'{e}n waves by multi-dimensional effects that are not included in our simulations.

\end{enumerate}



\acknowledgments

Hinode is a Japanese mission developed and launched by ISAS/JAXA, with NAOJ as domestic partner and NASA and STFC (UK) as international partners. 
It is operated by these agencies in co-operation with ESA and NSC (Norway).
Hinode is a Japanese mission developed and launched by ISAS/JAXA, collaborating with NAOJ as a domestic partner, NASA and STFC (UK) as international partners. 
Scientific operation of the Hinode mission is conducted by the Hinode science team organized at ISAS/JAXA. 
This team mainly consists of scientists from institutes in the partner countries. 
Support for the post-launch operation is provided by JAXA and NAOJ (Japan), STFC (U.K.), NASA, ESA, and NSC (Norway).
This work was (partly) carried out at the NAOJ Hinode Science Center, which is supported by the Grant-in-Aid for Creative Scientific Research "The Basic Study 
of Space Weather Prediction" from MEXT, Japan (Head Investigator: K. Shibata), generous donations from Sun Microsystems, and NAOJ internal funding.
Takuma Matsumoto is supported by the Research Fellowship from the Japan Society for the Promotion of Science for Young Scientists.

\clearpage

\begin{table}
\tabletypesize{\scriptsize}
\rotate
\begin{center}
\caption{Horizontal velocity spectrum in quiet region. \label{tbl-1}}
\begin{tabular}{cccccccc}
\tableline\tableline
\# of region & Start Time & Duration & Mean Cadence &  $\sqrt{<V_{\phi}^2>}$ & $\nu _b$ & $\alpha_L$ & $\alpha_H$\\
 & & (min) & (second) & (km s$^{-1}$) & (mHz) & & \\
\tableline
  1&31-Oct-2006 11:10:31& 79&30.0& 1.10& 4.7&-0.6&-2.4\\
  2&02-Nov-2006 23:40:30& 99& 32.0& 1.07& 5.0&-0.7&-2.6\\
  3&29-Dec-2006 16:50:30&308&30.0& 1.06& 4.3&-0.6&-2.1\\
  4&30-Dec-2006 04:12:29&272&30.0& 1.07& 4.4&-0.6&-2.1\\
  5&15-Feb-2007 08:14:50& 75&34.2& 1.07& 4.8&-0.6&-2.4\\
  6&17-Feb-2007 12:35:02&174&33.6& 1.11& 5.1&-0.7&-2.6\\
  7&19-Feb-2007 18:19:02&141&33.6& 1.13& 5.0&-0.7&-2.5\\
  8&27-Feb-2007 03:34:32&119&33.6& 1.07& 4.7&-0.6&-2.4\\
  9&03-Mar-2007 07:35:31&130&33.6& 1.10& 4.8&-0.6&-2.5\\
 10&09-Mar-2007 04:38:08& 81& 32.0& 1.10& 5.0&-0.7&-2.5\\
 11&18-Mar-2007 07:56:04& 81& 32.0& 1.06& 5.0&-0.7&-2.6\\
 12&06-Apr-2007 18:24:02&345&30.0& 1.08& 4.5&-0.6&-2.1\\
 13&14-Apr-2007 18:13:31&345&30.0& 1.10& 4.5&-0.6&-2.1\\
 14&29-Dec-2007 13:35:34&278&30.0& 1.13& 4.4&-0.6&-2.1\\
\tableline
\end{tabular}
\end{center}
\end{table}

\end{document}